\documentclass[
aps,prd,
12pt,%10pt
nopreprintnumbers,
showpacs,
eqsecnum,
nofootinbib
]{revtex4-1}

\usepackage{graphicx}

\def\v0{{\bf 0}}

\begin{document}

\title{
Critical Cosmology in Higher Order Gravity}
\author{Nahomi Kan}\email[]{kan@yamaguchi-jc.ac.jp}
\affiliation{
Yamaguchi Junior College,
Hofu-shi, Yamaguchi 747--1232, Japan}
\author{Koichiro Kobayashi}\email[]{m004wa@yamaguchi-u.ac.jp}
\author{Kiyoshi Shiraishi}\email[]{shiraish@yamaguchi-u.ac.jp}
\affiliation{
Yamaguchi University,
Yamaguchi-shi, Yamaguchi 753--8512, Japan}
\date{\today}
%\date{}

\begin{abstract}
We construct the higher order terms of curvatures in Lagrangians
of the scale factor for the Friedmann-Lema\^itre-Robertson-Walker
universe, which are linear in the second derivative of the scale factor
with respect to cosmic time. It is shown that they are composed from the
Lovelock tensors at the first step; iterative construction yields
arbitrarily high order terms.
The relation to the former work on higher order gravity is discussed.
Despite the absence of scalar degrees of freedom in cosmological models
which come from our Lagrangian, it is shown that an
inflationary behavior of the scale factor can be found.
The application to the thick brane solutions is also studied.
\end{abstract}

%\preprint{}


\maketitle

%%%%%%%%%%%%%%%%%%%%%%%%%%%%%%%%%%%%%%%%%%%%%%%%%%%%%%%%%%%%%%%%%%%%%%%%%%%
%Introduction
%%%%%%%%%%%%%%%%%%%%%%%%%%%%%%%%%%%%%%%%%%%%%%%%%%%%%%%%%%%%%%%%%%%%%%%%%%%
%%%%%%%%%%%%%%%%%%%%%%%%%%%%%%%%%%%%%%%%%%%%%%%%%%%%%%%%%%%%%%%%%%%%%%%%%%%
\section{Introduction}
%%%%%%%%%%%%%%%%%%%%%%%%%%%%%%%%%%%%%%%%%%%%%%%%%%%%%%%%%%%%%%%%%%%%%%%%%%%
%%%%%%%%%%%%%%%%%%%%%%%%%%%%%%%%%%%%%%%%%%%%%%%%%%%%%%%%%%%%%%%%%%%%%%%%%%%
It is well known that higher-derivative gravity has a scalar degree of
freedom in general \cite{Schmidt,Whitt,HL}. In cosmological models of 
higher-derivative gravity, the scalar mode is expected to play an
important role \cite{fR1,fR2,fR3}.
On the other hand, some cases are also known that higher order terms in
curvatures for a gravitational action do not affect cosmological
development of a scale factor.
For example, it is known that terms which consist of contraction
of Weyl tensors in a gravitational 
Lagrangian do not change evolutional equations for a scale factor in a
model with homogeneous and isotropic space.
The other special combinations of curvatures are known. 
In the specific dimension, the Euler form as a
Lagrangian does not produce the dynamics of gravity at all, because the
action becomes a topological quantity in such a case. 

The
dimensionally continued Euler densities have also been studied
\cite{Lovelock,JTWheeler,Aragone,MH,WP,Zumino,Zwi},
 because
of their relation to the effective Lagrangian of string theory, and are
found to give no scalar mode since the second derivative of the metric
disappears in the action if we perform integration by parts. The absence
of scalar modes is interesting for studying black holes in the theory,
because the scalar modes lead to singularities, in general, which avoid
expected horizons. 

In recent years, it turned out
that there is a special case where a scalar mode disappears in
higher-derivative gravity. Originally, this fact was found in research of
a three-dimensional theory of massive gravity \cite{NMG} and an extended
version in four dimensions was proposed \cite{LP}. The authors of those
papers intended to study the renormalizability and unitarity of
gravitation theory in a maximally-symmetric spacetime. Thus, the absence
of a massive scalar mode is at least a necessary condition of such
theories referred as critical gravity.%
\footnote{In our analysis, we do not care for the other critical value
for the cosmological constant, etc.}
 Until now, however, only the cases with
curvature tensors of a limited number have been investigated in higher
dimensions \cite{LPP,SGT}.  We are interested in the higher order
theory of gravitation in which a scalar mode does not appear in a general
higher dimensions. 

In the present paper, we generalize the structure of the
Lagrangian of critical gravity, to models with higher order terms in
curvature tensors in higher dimensions. We show that such extensions can
be attained by use of the Lovelock tensors. In order to offer a
systematic way to construct the required higher order term, we
take an explanatory approach by assuming the
Friedmann-Lema\^itre-Robertson-Walker (FLRW) metric. In this approach,
the absence of second derivatives of the scale factor from the Lagrangian
with appropriate total derivatives is considered as a necessary condition
for disappearing scalar modes. 

%we owe the
%specific metric in the analysis without a loss of generality. 

It should be noted that the combination defined in
$D$-dimensional spacetime
\begin{equation}
R_{\mu\nu}R^{\mu\nu}-\frac{D}{4(D-1)}R^2\,,
\end{equation}
is used in critical
gravities in three and four dimensions \cite{NMG,LP}. 
This term can be considered as a trace of multiplication of the Einstein
tensor and a linear combination of the Einstein tensor and its trace
part:
\begin{equation}
\left(R_{\mu\nu}-\frac{1}{2}R\, g_{\mu\nu}\right)
\left[R^{\mu\nu}-\frac{1}{2(D-1)}R\, g^{\mu\nu}\right]
=
G_{\mu\nu}\left(G^{\mu\nu}-\frac{1}{D-1}G^\lambda_\lambda\,
g^{\mu\nu}\right)\,,
\end{equation}
where $G_{\mu\nu}=R_{\mu\nu}-\frac{1}{2}R\, g_{\mu\nu}$ is the Einstein
tensor. Incidentally, $R^{\mu\nu}-\frac{1}{2(D-1)}R\, g^{\mu\nu}$
is known as the Schouten tensor up to the factor $(D-2)^{-1}$.
Now, if the FLRW metric is assumed, the time-time component of the
Einstein tensor does not include the second time-derivative of the scale
factor. The Schouten tensor appearing the above term is made for its
spatial component to have no second time-derivative of the scale
factor.  Therefore, the trace of the product of two tensors is linear in
the second time-derivative of the scale factor, and if a
surface term is suitably assigned, the Lagrangian is
expressed only with the scale factor and its first time-derivative.
Thus, additional scalar modes do not appear. From this observation, we
find that it
can be extended by using the Lovelock tensor instead of the Einstein
tensor when the dimension of spacetime is higher.
In the present paper, we do not analyze the massive tensor modes in our
models. Thus, the genuine criticality as quantum gravity is left for
future works.

The FLRW geometry is known to be conformally flat \cite{Ibison}, i.e.,
the Weyl tensor for the FLRW cosmological metrics vanishes.
An extension of Lovelock gravity for conformally-flat geometry was
considered by Meissner and Olechowski \cite{MO}.
They showed that the extension is possible for a $(R)^n$ term in $D$
dimensions provided $n<D$, whereas the Lovelock gravity has the
$(R)^n$ term at most $2n<D$ (where $(R)$ denotes a general curvature
tensor). 
Oliva and Ray also construct higher-derivative gravity with the second
order equation utilizing the Weyl and Riemann tensors \cite{OR}. Their
Lagrangian involves up to
$(R)^{n}$ term with $2n<D$ because of the use of the generalized Kronecker
delta as in the case of the Lovelock gravity.

In the present paper, we show that it is
possible to continue the higher-curvature terms beyond the number of
dimensions.

The outline of this paper is as follows.
In \S2, we construct the candidate Lagrangian for 
cosmological models without scalar modes in the tensorial form as the
first step.  The confirmation of the property of the model Lagrangian is
performed  in
\S3, substituting the FLRW metric.  In \S4, we show that an
extension to more higher order terms in curvatures can be obtained.
Using the higher order Lagrangian so far obtained, we propose (toy)
models for the scale factor with inflationary behavior is shown in \S5. 
In \S6, the application of our Lagrangian to constructing domain wall
solutions is studied.  In the last section, we offer some concluding
remarks.

%%%%%%%%%%%%%%%%%%%%%%%%%%%%%%%%%%%%%%%%%%%%%%%%%%%%%%%%%%%%%%%%%%%%%%%%%%%
%%%%%%%%%%%%%%%%%%%%%%%%%%%%%%%%%%%%%%%%%%%%%%%%%%%%%%%%%%%%%%%%%%%%%%%%%%%
\section{Lovelock tensors and generalization of the higher order term
in critical gravity}
%%%%%%%%%%%%%%%%%%%%%%%%%%%%%%%%%%%%%%%%%%%%%%%%%%%%%%%%%%%%%%%%%%%%%%%%%%%
%%%%%%%%%%%%%%%%%%%%%%%%%%%%%%%%%%%%%%%%%%%%%%%%%%%%%%%%%%%%%%%%%%%%%%%%%%%
In this section, we construct the higher order term in curvatures by
generalizing that of critical gravity.
We will verify the absence of scalar modes in cosmological models with the
terms in the next section. First, we introduce the
dimensionally continued Euler density
\begin{equation}
{L^{(n)}}=2^{-n}
\delta^{\sigma_1\tau_1\cdots\sigma_n\tau_n}_{\lambda_1\rho_1\cdots\lambda_n\rho_n}
R^{\lambda_1\rho_1}{}_{\sigma_1\tau_1}\cdots
R^{\lambda_n\rho_n}{}_{\sigma_n\tau_n}\,,
\end{equation}
where the generalized Kronecker delta is defined as
\begin{equation}
\delta^{\mu_1\mu_2\cdots\mu_p}_{\nu_1\nu_2\cdots\nu_p}
=\left|\begin{array}{cccc}
\delta^{\mu_1}_{\nu_1} & \delta^{\mu_1}_{\nu_2} & \cdots &
\delta^{\mu_1}_{\nu_p} \\
\delta^{\mu_2}_{\nu_1} & \delta^{\mu_2}_{\nu_2} & \cdots &
\delta^{\mu_2}_{\nu_p} \\
\vdots & \vdots & \ddots & \vdots \\
\delta^{\mu_p}_{\nu_1} & \delta^{\mu_p}_{\nu_2} & \cdots &
\delta^{\mu_p}_{\nu_p} 
\end{array}\right|\,.
\end{equation}
The
dimensionally continued Euler density $L^{(n)}$ consists of $n$-th order
in the curvature tensors ($(R)^n$). For example, for $n=1$, we find the
Einstein-Hilbert term
\begin{equation}
{L^{(1)}}=R\,,
\end{equation}
and for $n=2$, we find the Gauss-Bonnet term
\begin{equation}
{L^{(2)}}=R_{\mu\nu\rho\sigma}R^{\mu\nu\rho\sigma}-4
R_{\mu\nu}R^{\mu\nu}+R^2\,,
\end{equation}
as is well known.

Next, we consider the Lovelock tensor \cite{Lovelock}.
The Lovelock tensor is a generalization of the Einstein tensor,
and defined as,
\begin{equation}
{G^{(n)}}^{\mu}_{\nu}\equiv
\frac{1}{\sqrt{-g}}\frac{\delta(\int d^Dx\sqrt{-g}L^{(n)})}{\delta
g^{\rho\nu}}g^{\mu\rho}
=-2^{-(n+1)}
\delta^{\mu\sigma_1\tau_1\cdots\sigma_n\tau_n}_{\nu\lambda_1\rho_1\cdots\lambda_n\rho_n}
R^{\lambda_1\rho_1}{}_{\sigma_1\tau_1}\cdots
R^{\lambda_n\rho_n}{}_{\sigma_n\tau_n}\,.
\end{equation}
This is a symmetric tensor of $n$-th order in the curvature tensors. For
example, for $n=1$, we find
\begin{equation}
G_{\mu\nu}^{(1)}=R_{\mu\nu}-\frac{1}{2}Rg_{\mu\nu}\,,
\end{equation}
and this is known as the Einstein tensor especially.
For $n=2$, we find
\begin{equation}
G_{\mu\nu}^{(2)}=2(R_{\mu\rho\sigma\tau}R_{\nu}{}^{\rho\sigma\tau}-2
R_{\mu\rho\nu\sigma}R^{\rho\sigma}-2
R_{\mu\rho}R_{\nu}^{\rho}+RR_{\mu\nu})-\frac{1}{2}{L^{(2)}}
g_{\mu\nu}\,.
\end{equation}
It should be noted that
\begin{equation}
G^{(n)}\equiv{G^{(n)}}_\lambda^\lambda=\frac{2n-D}{2}\,L^{(n)}\,,
\end{equation}
where $D$ denotes the dimension of spacetime.

Here, we construct the new combination of the Lovelock tensor and the
metric multiplicated by the trace of the Lovelock tensor.
That is,
\begin{equation}
S^{\mu\nu}_{(n)}\equiv{G^{(n)}}^{\mu\nu}-\frac{1}{D-1}
G^{(n)}g^{\mu\nu}\,.
\end{equation}
It is worth noting that
\begin{equation}
G^{(n)}_{\mu\nu}=(g_{\mu\rho}g_{\nu\sigma}-
g_{\mu\nu}g_{\rho\sigma})S^{\rho\sigma}_{(n)}\,.
\end{equation}
For example, for $n=1$, we obtain
\begin{equation}
S^{\mu\nu}_{(1)}=R^{\mu\nu}-\frac{1}{2(D-1)}Rg^{\mu\nu}\,,
\end{equation}
which is proportional to the Schouten tensor.
Therefore we obtain the following combination
\begin{equation}
G^{(1)}_{\mu\nu}S^{\mu\nu}_{(1)}=R_{\mu\nu}R^{\mu\nu}-
\frac{D}{4(D-1)}R^2\,,
\end{equation}
which appears in critical gravities \cite{NMG,LP}.

Now, we find that the natural generalization of this is given by
\begin{equation}
{G^{(n)}}_{\mu\nu}{S_{(n')}}^{\mu\nu}=
{G^{(n)}}_{\mu\nu}{G^{(n')}}^{\mu\nu}-\frac{1}{D-1}
G^{(n)}G^{(n')}\,.
\end{equation}
It is worth pointing out that the expression is symmetric against the
exchange of
$n$ and
$n'$.

In the next section, we confirm that this combination is suitable for 
an extension of critical gravity in higher dimensions, by utilizing the
FLRW metric.

%%%%%%%%%%%%%%%%%%%%%%%%%%%%%%%%%%%%%%%%%%%%%%%%%%%%%%%%%%%%%%%%%%%%%%%%%%%
%%%%%%%%%%%%%%%%%%%%%%%%%%%%%%%%%%%%%%%%%%%%%%%%%%%%%%%%%%%%%%%%%%%%%%%%%%%
\section{higher order term for FLRW metric}
%%%%%%%%%%%%%%%%%%%%%%%%%%%%%%%%%%%%%%%%%%%%%%%%%%%%%%%%%%%%%%%%%%%%%%%%%%%
%%%%%%%%%%%%%%%%%%%%%%%%%%%%%%%%%%%%%%%%%%%%%%%%%%%%%%%%%%%%%%%%%%%%%%%%%%%
We consider the following FLRW metric in $D$ dimensions:
\begin{equation}
ds^2=-dt^2+a(t)^2d\Omega^2_{D-1}\,,
\end{equation}
where $a(t)$ is the scale factor and $d\Omega^2_{D-1}$ denotes the line
element of a maximally symmetric space of $(D-1)$-dimensions, whose scalar
curvature is normalized to $(D-1)(D-2)k$
with $k=1, 0, -1$.

First in this section, we examine the Lovelock tensors of $(R)^n$.
By explicit calculation of curvatures, we find, for $n=1$,
\begin{eqnarray}
{G^{(1)}}_0^0&=&-\frac{(D-1)(D-2)}{2}\frac{\dot{a}^2+k}{a^2}\,,\\
{G^{(1)}}_j^i&=&-\left[(D-2)\frac{\ddot{a}}{a}+\frac{(D-2)(D-3)}{2}\frac{\dot{a}^2+k}{a^2}\right]
\delta^i_j\,,
\end{eqnarray}
where $\dot{a}=\frac{da}{dt}$ and  $\ddot{a}=\frac{d^2a}{dt^2}$.
Here and hereafter, we use the suffixes denoting the spatial dimensions,
$i, j=1, 2,\dots ,D-1$. Also, for $n=2$, we obtain
\begin{eqnarray}
{G^{(2)}}^0_0&=&-\frac{(D-1)(D-2)(D-3)(D-4)}{2}
\left(\frac{\dot{a}^2+k}{a^2}\right)^2\,,\\
{G^{(2)}}^i_j&=&-\left[2(D-2)(D-3)(D-4)\frac{\ddot{a}}{a}
\frac{\dot{a}^2+k}{a^2}\right.\nonumber \\
& &\qquad+\left.
\frac{(D-2)(D-3)(D-4)(D-5)}{2}\left(\frac{\dot{a}^2+k}{a^2}\right)^2\right]
\delta^i_j\,.
\end{eqnarray}
%For $n=3$, we obtain 
%\begin{eqnarray}
%{G^{(3)}}^0_0&=&-\frac{(D-1)(D-2)(D-3)(D-4)(D-5)(D-6)}{2}
%\left(\frac{\dot{a}^2+k}{a^2}\right)^3\,,\\
%{G^{(3)}}^i_j&=&-\left[3(D-2)(D-3)(D-4)(D-5)(D-6)
%\frac{\ddot{a}}{a}\left(
%\frac{\dot{a}^2+k}{a^2}\right)^2\right.\nonumber \\
%& &\qquad+\left.
%\frac{(D-2)(D-3)(D-4)(D-5)(D-6)(D-7)}{2}
%\left(\frac{\dot{a}^2+k}{a^2}\right)^3\right]
%\delta^i_j\,.
%\end{eqnarray}

By the combinatorial property of the generalized Kronecker delta, we can
find the Lovelock tensor for a general $n$ as follows:
\begin{eqnarray}
{G^{(n)}}^0_0&=&-\frac{(D-1)(D-2)\cdots(D-2n)}{2}
\left(\frac{\dot{a}^2+k}{a^2}\right)^n\,,\\
{G^{(n)}}^i_j&=&-\left[n(D-2)(D-3)\cdots(D-2n)\frac{\ddot{a}}{a}\left(
\frac{\dot{a}^2+k}{a^2}\right)^{n-1}\right.\nonumber \\
& &\qquad+\left.
\frac{(D-2)(D-3)\cdots(D-2n-1)}{2}\left(\frac{\dot{a}^2+k}{a^2}\right)^n\right]
\delta^i_j\,,
\end{eqnarray}
and ${G^{(n)}}^0_j={G^{(n)}}^i_0=0$.
It should be noted that the $00$ component of the Lovelock tensor does
not include $\ddot{a}$.

Next, we calculate the generalized Schouten tensor ${S_{(n)}}^\mu_\nu$ for
the FLRW metric. Because the trace of the Lovelock tensor is given as
\begin{eqnarray}
\frac{1}{D-1}{G^{(n)}}&=&-\left[n(D-2)(D-3)\cdots(D-2n)\frac{\ddot{a}}{a}\left(
\frac{\dot{a}^2+k}{a^2}\right)^{n-1}\right.\nonumber \\
& &\qquad+\left.(D-2n)
\frac{(D-2)(D-3)\cdots(D-2n)}{2}\left(\frac{\dot{a}^2+k}{a^2}\right)^n\right]
\,,
\end{eqnarray}
we find
\begin{eqnarray}
{S_{(n)}}^0_0&=&n(D-2)(D-3)\cdots(D-2n)\frac{\ddot{a}}{a}\left(
\frac{\dot{a}^2+k}{a^2}\right)^{n-1}\nonumber \\
& &\qquad-(2n-1)
\frac{(D-2)(D-3)\cdots(D-2n)}{2}\left(\frac{\dot{a}^2+k}{a^2}\right)^n
\,,\\
{S_{(n)}}^i_j&=&\frac{(D-2)(D-3)\cdots(D-2n)}{2}\left(\frac{\dot{a}^2+k}{a^2}\right)^n
\delta^i_j\,.
\end{eqnarray}
By construction, $\ddot{a}$ is absent in ${S_{(n)}}^i_j$.

Now, we consider the combined term
${G^{(n)}}_{\mu\nu}{S_{(n')}}^{\mu\nu}$. We obtain
\begin{eqnarray}
{G^{(n)}}_{\mu\nu}{S_{(n')}}^{\mu\nu}&=&-\frac{1}{4}(D-1)[(D-2)(D-3)
\cdots(D-2n)][(D-2)(D-3)
\cdots(D-2n')]\nonumber \\
& &\times\left[2(n+n')\frac{\ddot{a}}{a}\left(
\frac{\dot{a}^2+k}{a^2}\right)^{n+n'-1}+(D-2(n+n'))
\left(\frac{\dot{a}^2+k}{a^2}\right)^{n+n'}\right]
\,.
\end{eqnarray}
Because this combination is apparently linear in $\ddot{a}$,
the action including this term can be expressed as the functional of $a$
and $\dot{a}$, by means of integrations by part.
Therefore, we realize that there is no scalar mode in a cosmological
setting.

The combination $L^{(n)(n')}\equiv
-4{G^{(n)}}_{\mu\nu}{S_{(n')}}^{\mu\nu}$ can exist for limited numbers
$(n, n')$, which depends on $D$. For the exchange symmetry, we assume
$n\le n'$. We find that $L^{(n)(n')}$ is non-zero when
$(n, n')=(1,1)$ for $D=3, 4$,
$(n, n')=(1,1), (1,2), (2,2)$ for $D=5, 6$ and
$(n, n')=(1,1), (1,2), (1,3), (2,2), (2,3), (3,3)$ for $D=7, 8$,
etc.
Therefore,
we have new higher-derivative terms including $(R)^m$ with $m\le D-1$
for odd $D$ and $m\le D-2$ for even $D$.

Up to now, we become aware of a similarity to the Lovelock Lagrangian.
It takes the form for the FLRW metric as follows:
\begin{eqnarray}
{L^{(m)}}&=&(D-1)(D-2)
\cdots(D-2m+1)\nonumber \\
& &\times\left[2m\frac{\ddot{a}}{a}\left(
\frac{\dot{a}^2+k}{a^2}\right)^{m-1}+(D-2m)
\left(\frac{\dot{a}^2+k}{a^2}\right)^{m}\right]
\,.
\end{eqnarray}
The equivalence up to the overall constant is obvious, that is
$L^{(n)(n')}\propto L^{(n+n')}$.
Incidentally, we can consider $L^{(0)(n)}$ as the Lovelock
Lagrangian $L^{(n)}$.
We note, however, that the Lovelock Lagrangian $L^{(m)}$ has its meaning
as a part of Lagrangian when $m=1$ for~$D=3, 4$, $m=1, 2$ for $D=5, 6$
and $m=1, 2, 3$ for $D=7, 8$, etc., because $L^{(m)}$ becomes a
total derivative for $D=2m$.

Before closing this section, discussion on  relation to the work of
Meissner and Olechowski
\cite{MO} is in order.
Their approach is equivalent
 to considering the Lagrangian constructed
from the Lovelock Lagrangian in which the Riemann tensor is expressed by
the Schouten tensor under the assumption of vanishing Weyl tensor.%
\footnote{Strictly speaking, they replace Riemann tensor by the
Kulkarni-Nomizu product of the Schouten tensor with the metric in the
Lovelock tensor.} 
They used the
combinatorial property of the generalized Kronecker delta for extension to
higher order in curvatures. If the FLRW metric is substituted, we find
that the Lagrangian of Meissner and Olechowski of order of $(R)^m$
coincides with our Lagrangian
$L^{(n)(n')}$, where $n+n'=m$, up to the constant factor.
The allowed spacetime dimension is the same for odd $m$, $m<D$.
Therefore, our Lagrangian and theirs are almost equivalent. The variety
with respect to two integers in $L^{(n)(n')}$ is due to the use 
of Riemann tensors as well as scalar curvatures and Ricci tensors in
our approach. It is notable that differences may occur if we consider the
black hole or  non-conformally flat solutions in the theory governed
by the Lagrangians of higher order terms.

Later, we show that the restriction by the dimensions can be overcome. 
To exhibit the discussion on the subject,
we examine the cosmological action in the present model again in the next
section.

%%%%%%%%%%%%%%%%%%%%%%%%%%%%%%%%%%%%%%%%%%%%%%%%%%%%%%%%%%%%%%%%%%%%%%%%%%%
%%%%%%%%%%%%%%%%%%%%%%%%%%%%%%%%%%%%%%%%%%%%%%%%%%%%%%%%%%%%%%%%%%%%%%%%%%%
\section{further extension to higher order in curvature (especially for $m
> D$)}
%%%%%%%%%%%%%%%%%%%%%%%%%%%%%%%%%%%%%%%%%%%%%%%%%%%%%%%%%%%%%%%%%%%%%%%%%%%
%%%%%%%%%%%%%%%%%%%%%%%%%%%%%%%%%%%%%%%%%%%%%%%%%%%%%%%%%%%%%%%%%%%%%%%%%%%
We consider the action for
$m\ge 2$:
\begin{equation}
S^{(m)}=\int d^Dx \sqrt{-g}\sum_{n=0}^{m}\alpha_{(n)(m-n)}L^{(n)(m-n)}\,,
\end{equation}
with arbitrary coefficients $\alpha_{(n)(m-n)}$.
Here, we regard $L^{(0)(n)}$ as $L^{(n)}$.

Then, the action for the scale factor $a(t)$ can be read as
\begin{eqnarray}
S^{(m)}\propto \beta_{m}\int dt\,
a^{D-1}\left[2m\frac{\ddot{a}}{a}\left(
\frac{\dot{a}^2+k}{a^2}\right)^{m-1}+(D-2m)
\left(\frac{\dot{a}^2+k}{a^2}\right)^{m}\right]
\,,
\end{eqnarray}
where
\begin{equation}
\beta_{m}=(D-1)\alpha_{(n)(m-n)}\frac{(D-2)!}{(D-2n-1)!}
\frac{(D-2)!}{(D-2(m-n)-1)!}\,.
\end{equation}

Using the expansion
\begin{equation}
(\dot{a}^2+k)^{m-1}\ddot{a}=\sum_{\ell=0}^{m-1}
\left(\begin{array}{c}
m-1 \\
\ell
\end{array}
\right)k^{m-1-\ell}\dot{a}^{2\ell}\ddot{a}
=\sum_{\ell=0}^{m-1}\frac{k^{m-1-\ell}}{2\ell+1}
\left(\begin{array}{c}
m-1 \\
\ell
\end{array}
\right)\frac{d}{dt}\dot{a}^{2\ell+1}
\,,
\end{equation}
we rewrite the action, after partial integration, as
\begin{eqnarray}
I^{(m)}[a, \dot{a}]\equiv(D-2m)\beta_{m}\int dt\,
a^{D-1}\left[-2m\sum_{\ell=0}^{m-1}\frac{k^{m-1-\ell}}{2\ell+1}
\left(\begin{array}{c}
m-1 \\
\ell
\end{array}
\right)\frac{\dot{a}^{2\ell+2}}{a^{2m}}+
\left(\frac{\dot{a}^2+k}{a^2}\right)^{m}\right]
\,.
\label{45}
\end{eqnarray}

Especially for $k=0$, we find a simple action
\begin{eqnarray}
I^{(m)}=(D-2m)\beta_{m}\int dt\,
a^{D-1}\left[-\frac{1}{2m-1}
\left(\frac{\dot{a}^2}{a^2}\right)^{m}\right]
\,.
\end{eqnarray}

Now, we define the Lagrangian ${\rm L}^{(m)}(a, \dot{a})$ for the scale
factor
$a(t)$ by
\begin{equation}
I^{(m)}=\int dt\,{\rm L}^{(m)}(a,\dot{a})\,.
\end{equation}
Then, we find the equation of motion for the scale factor
\begin{eqnarray}
& &\frac{d}{dt}\frac{\partial {\rm L}^{(m)}}{\partial \dot{a}}-
\frac{\partial {\rm L}^{(m)}}{\partial a}\nonumber \\
&=&-(D-2m)\beta_{m}\,
a^{D-2}\left[2m\frac{\ddot{a}}{a}\left(\frac{\dot{a}^2+k}{a^2}\right)^{m-1}
+(D-2m-1)
\left(\frac{\dot{a}^2+k}{a^2}\right)^{m}\right]
\,.
\label{ii}
\end{eqnarray}

The Hamiltonian constraint is regarded as the variation equation
$\frac{\partial {\rm L}}{\partial N}=0$, where $N$ is the lapse function
defined as $dt=N(t')dt'$. We set $N=1$ after the manipulation.
We now find the Hamiltonian constraint equation
\begin{equation}
\left.\frac{\partial{\rm L}^{(m)}}{\partial N}\right|_{N=1}=
(D-2m)\beta_{m}\,
a^{D-1}\left(\frac{\dot{a}^2+k}{a^2}\right)^{m}\,.
\label{00}
\end{equation}

It is known that the variation of the lapse function corresponds to the
variation of $g_{00}$ and the variation of the scale factor corresponds to
the variation of $g_{ii}$, up to certain factors. We can find, from
(\ref{00}) and (\ref{ii}), the following generalized Lovelock tensor as
the variation of the action with respect to the metric:
\begin{eqnarray}
{\bar{G}^{(m)}}{}^0_0&=&-\frac{(D-2m)\beta_{m}}{2}
\left(\frac{\dot{a}^2+k}{a^2}\right)^m\,,\\
{\bar{G}^{(m)}}{}^i_j&=&-\frac{(D-2m)\beta_{m}}{2(D-1)}\left[2m\frac{\ddot{a}}{a}\left(
\frac{\dot{a}^2+k}{a^2}\right)^{m-1}+
(D-2m-1)\left(\frac{\dot{a}^2+k}{a^2}\right)^m\right]
\delta^i_j\,.
\end{eqnarray}
The generalized Lovelock tensor has a tensorial form, which is
proportional to
\begin{equation}
\bar{G}^{(m)}_{\mu\nu}\propto\frac{1}{\sqrt{-g}}\frac{\delta
S^{(m)}}{\delta g^{\mu\nu}}\,,
\end{equation}
as seen from the
construction, but with arbitrary coefficients in the definition of the
action $S^{(m)}$. In spite of the arbitrariness, the functional
form of the Lagrangian is unambiguous if the conformally flat metric is
substituted.

In the similar manner, the corresponding generalized Schouten tensor is
defined as in the previous section,
\begin{equation}
\bar{S}^{\mu\nu}_{(m)}\equiv \bar{G}^{(m)}{}^{\mu\nu}-\frac{1}{D-1}
\bar{G}^{(m)}g^{\mu\nu}\,,
\end{equation}
with $\bar{G}^{(m)}\equiv g_{\mu\nu}\bar{G}^{(m)}{}^{\mu\nu}$.
Then we find
\begin{eqnarray}
{\bar{S}_{(m)}}{}_0^0&=&\frac{(D-2m)\beta_{m}}{2(D-1)}\left[2m\frac{\ddot{a}}{a}\left(
\frac{\dot{a}^2+k}{a^2}\right)^{m-1}-
(2m-1)\left(\frac{\dot{a}^2+k}{a^2}\right)^m\right]
\,,\\
{\bar{S}_{(m)}}{}_j^i&=&\frac{(D-2m)\beta_{m}}{2(D-1)}
\left(\frac{\dot{a}^2+k}{a^2}\right)^m\delta^i_j\,.
\end{eqnarray}
Then, the trace of the product of these tensors is found to be
\begin{eqnarray}
& &L^{(m)(m')}=-4\bar{G}^{(m)}_{\mu\nu}\bar{S}_{(m')}^{\mu\nu}\nonumber \\
&=&
\frac{(D-2m)(D-2m')\beta_{m}\beta_{m'}}{D-1}\,
\left[2(m+m')\frac{\ddot{a}}{a}
+\{D-2(m+m')\}\frac{\dot{a}^2+k}{a^2}\right]
\left(\frac{\dot{a}^2+k}{a^2}\right)^{m+m'-1}
\,.
\end{eqnarray}

Again, we obtain the same functional form as those
constructed from the Lovelock tensors in the previous section.
More iterative operations yield the more higher order terms. For this
time, however, no limitation on the relation between the number of
dimensions
$D$ and the order of curvatures
$\ell\equiv m+m'$, except for the case with $2\ell=D$ (the Lagrangian
becomes a total derivative in this case for the conformally flat
spacetime). Although we do not exhibit an explicit tensorial form of the
higher order term in mass dimensions, obviously it can be expressed as the
 combination of curvature tensors and their covariant derivatives. 

Unfortunately, only for $D=4$, our iterative method cannot give a
$(R)^3$ term, since the $(R)^2$ term becomes a total derivative for the
conformally flat metric. We define the expression for
$L^{(3)}$ by the manner followed by Meissner and Olechowski as
\begin{equation}
L^{(3)}\propto\delta^{\mu_1\mu_2\mu_3}_{\nu_1\nu_2\nu_3}
{S_{(1)}}^{\nu_1}_{\mu_1}{S_{(1)}}^{\nu_2}_{\mu_2}{S_{(1)}}^{\nu_3}_{\mu_3}
\,.
\end{equation}
Incidentally, it is obvious that $L^{(3)}$ is at most linear in $\ddot{a}$
if the FLRW metric is substituted, because only $S_{(1)}{}^0_0$ is the
component of the Schouten tensor which is linear in $\ddot{a}$.  Then,
the desired terms of all the order in curvatures and derivatives in any
dimensions can be created by our iterative method.

Another way to cross the dimensional limitation, which is inspired
by the work of Meissner and Olechowski \cite{MO}, is also found.
In the manner of their paper \cite{MO}, the following Lagrangian (modulo
volume form) is proposed:
\begin{equation}
\delta^{\mu_1\cdots\mu_n}_{\nu_1\cdots\nu_n}
{S_{(1)}}^{\nu_1}_{\mu_1}\cdots{S_{(1)}}^{\nu_n}_{\mu_n}
\,,
\end{equation}
where it is expressed by our present notation.
We become aware of an extension of it as
\begin{equation}
\delta^{\mu_1\cdots\mu_n}_{\nu_1\cdots\nu_n}
{\bar{S}_{(m_1)}}{}^{\nu_1}_{\mu_1}\cdots{\bar{S}_{(m_n)}}{}^{\nu_n}_{\mu_n}
\,,
\end{equation}
where $m_i ~(i=1,\dots, n)$ are arbitrary integers.
For the FLRW metric, this term turns out to be proportional to
\begin{equation}
2\ell\frac{\ddot{a}}{a}\left(\frac{\dot{a}^2+k}{a^2}\right)^{\ell-1}
+(D-2\ell)
\left(\frac{\dot{a}^2+k}{a^2}\right)^{\ell}
\,,
\end{equation}
where $\ell=\sum_{i=1}^n m_i$. This is also the same form of the candidate
Lagrangian.
Many equivalent combinations exist for higher order terms under
the vanishing-Weyl-tensor condition.

In the next section and after, 
we will turn our attention to apply the higher order Lagrangian
obtained here to  cosmological models .

%%%%%%%%%%%%%%%%%%%%%%%%%%%%%%%%%%%%%%%%%%%%%%%%%%%%%%%%%%%%%%%%%
%%%%%%%%%%%%%%%%%%%%%%%%%%%%%%%%%%%%%%%%%%%%%%%%%%%%%%%%%%%%%%%%%
\section{$f(H^2)$ cosmology}
%%%%%%%%%%%%%%%%%%%%%%%%%%%%%%%%%%%%%%%%%%%%%%%%%%%%%%%%%%%%%%%%%
%%%%%%%%%%%%%%%%%%%%%%%%%%%%%%%%%%%%%%%%%%%%%%%%%%%%%%%%%%%%%%%%%
In this section, we investigate the possible 
inflationary stage in evolution of the universe in the model with higher
order terms discussed in the previous sections.

Let us consider the general cosmological action  
\begin{equation}
{\rm
L}=a^{D-1}\sum_{\ell=0}^\infty\beta_\ell\left[2\ell\frac{\ddot{a}}{a}
\left(\frac{\dot{a}^2+k}{a^2}\right)^{\ell-1}
+(D-2\ell)
\left(\frac{\dot{a}^2+k}{a^2}\right)^{\ell}\right]
\,,
\label{action}
\end{equation}
with an appropriate total derivative term with respect to the time,
which removes the second derivative of the scale factor with respect to
time.

As usual, the energy momentum tensor of matter is taken as
\begin{equation}
T_{\mu\nu}=\frac{2}{\sqrt{-g}}\frac{\delta S_{matter}}{\delta
g^{\mu\nu}}\,,
\end{equation}
with the action for matter $S_{matter}$.
The non-zero components of the energy momentum tensor of matter is
assumed as
$T_0^0=-\rho$ and $T^i_j=p\,\delta^i_j$, where $\rho$ is the energy
density and $p$ is the pressure, for the FLRW universe. As is seen in the
previous sections, the equation of motion derived from the action
(\ref{action}) coincides with the linear combination of the 
components of the Lovelock tensors in functional form with the assumption
of the FLRW metric. Thus, the energy conservation
\begin{equation}
\dot{\rho}+(D-1)\frac{\dot{a}}{a}(\rho+p)=0\,,
\end{equation}
holds generally. This fact is due to the absence of the scalar mode and
the fact that we did not need to rescale the metric.

We here give a few
model
in four-dimensional spacetime with focus on the possibility of
an inflationary growth of the scale factor.
Furthermore, we take
$k=0$, i.e., assume the flat space.
Then, the action (\ref{action}) is equivalent to
\begin{equation}
{\rm
L}(a,
\dot{a})=a^{3}\sum_{\ell=0}^\infty\beta_\ell\left[-\frac{4-2\ell}{2\ell-1}
\left(\frac{\dot{a}^2}{a^2}\right)^{\ell}\right]
\,.
\end{equation}
The 00 component of the equation of motion reads
\begin{equation}
\sum_{\ell=0}^\infty\beta_\ell\frac{4-2\ell}{2}
\left(\frac{\dot{a}^2}{a^2}\right)^{\ell}=\rho
\,.
\end{equation}

If we can choose the coefficient $\beta_\ell$ freely,
almost arbitrary equations including $\rho$ and $H\equiv
\frac{\dot{a}}{a}$ can be made.
That is,
\begin{equation}
f(H^2)=\frac{8\pi G}{3}\rho
\,,
\end{equation}
where $f(x)$ is a function which can be expressed by a series and does
not include the $x^2$ term. Here the factor given in the right hand side
is chosen as for the similarity with the standard cosmology. We wish to
call the cosmology of the model `$f(H^2)$ cosmology'.%
\footnote{If $k\ne 0$, we obtain `$f(H^2+k/a^2)$ cosmology'.}

Now, we specify the function $f(x)$.
For example, let us consider
\begin{equation}
H^2\left(1-\frac{H^4}{M^4}\right)^{-1}=\frac{8\pi G}{3}\rho\,,
\end{equation}
with a typical mass scale $M$.
Solving the equation for $H^2$, we obtain
\begin{equation}
H^2=\frac{8\pi
G}{3}\rho\left[\frac{1}{2}+\sqrt{\frac{1}{4}+\left(\frac{8\pi
G\rho}{3M^2}\right)^2}\right]^{-1}\,.
\end{equation}

If the energy density is sufficiently low, such as $\rho\ll G^{-1}M^2$,
the relation in the standard cosmology is valid:
\begin{equation}
H^2\approx\frac{8\pi
G}{3}\rho\,.
\end{equation}
On the other hand, in the era of the high energy density such that
$\rho\gg G^{-1}M^2$, we find
\begin{equation}
H^2\approx M^2\,,
\end{equation}
and we obtain an approximate
de Sitter inflation.

Another model can be chosen, in which the 
correction is a monomial of the higher order term. That is,
\begin{equation}
H^2+M^2\left(\frac{H^2}{M^2}\right)^{m}=\frac{8\pi G}{3}\rho\qquad
(m\ge 3)\,.
\end{equation}
In the rapid expansion phase, $H^2\gg M^2$, the correction would be
dominant. Further, if we assume the dust matter, i.e., $\rho\propto
a^{-3}$, we obtain
\begin{equation}
a(t)\propto t^{\frac{2m}{3}}\,.
\end{equation}
This indicates the phase of the power-law inflation.

Even though we show special toy models here, we find that
the existence of higher order terms yields
the inflationary growth of the scale factor with ordinary matter,
and with no scalar mode and no redefinition of the metric.

%%%%%%%%%%%%%%%%%%%%%%%%%%%%%%%%%%%%%%%%%%%%%%%%%%%%%%%%%%%%%%%%%
%%%%%%%%%%%%%%%%%%%%%%%%%%%%%%%%%%%%%%%%%%%%%%%%%%%%%%%%%%%%%%%%%
\section{domain wall}
%%%%%%%%%%%%%%%%%%%%%%%%%%%%%%%%%%%%%%%%%%%%%%%%%%%%%%%%%%%%%%%%%
%%%%%%%%%%%%%%%%%%%%%%%%%%%%%%%%%%%%%%%%%%%%%%%%%%%%%%%%%%%%%%%%%
We shall apply the previous discussion on a conformally flat metric
to solutions for
domain walls (``thick branes'').
The $D$-dimensional metric suitable for a co-dimension one domain wall is
given by
\begin{equation}
ds^2=e^{2A(y)}(-dt^2+d{\bf x}^2)+dy^2\,,
\label{dw}
\end{equation}
where ${\bf x}=(x^1, x^2, \cdots x^{D-2})$.
In addition, we consider a neutral, minimally-coupled, self-interacting
scalar field as a classical matter field.
It is known that superpotential method \cite{ST,spd1,spd2,spd3,LCS2}
is available in this case to find BPS kink equations.
Let us investigate the case with higher order terms in a similar manner.

Substituting the metric (\ref{dw}), the gravitational Lagrangian becomes
\begin{equation}
{\rm
L}=e^{(D-1)A}\sum_{\ell=0}^\infty (-1)^\ell\beta_\ell
\left[2\ell A''\left(A'\right)^{2\ell-2}
+D
\left(A'\right)^{2\ell}\right]
\,,
\end{equation}
where the prime (${~}'$) denotes a derivative with respect to $y$.
The term in $A''$ can be removed by discarding a total divergence.
Thus we obtain
\begin{equation}
{\rm
L}(A, A')=e^{(D-1)A}\sum_{\ell=0}^\infty (-1)^\ell\beta_\ell
\left[-\frac{D-2\ell}{2\ell-1}
\left(A'\right)^{2\ell}\right]
\,.
\end{equation}
The action for the
real scalar field $\phi(y)$ for static branes is written by
\begin{equation}
{\rm
L}_S(A, \phi, \phi')=e^{(D-1)A}
\left[-\frac{1}{2}(\phi')^2-V(\phi)\right]
\,,
\end{equation}
Since the coefficients $\beta_\ell$ can be arbitrarily chosen,
we use an arbitrary function $F(x)$ to represent the general action.
Then, we can write the total action as
\begin{equation}
{\rm
L}_{total}(A, A', \phi, \phi')=e^{(D-1)A}
\left[F(A'^2)-\frac{1}{2}(\phi')^2-V(\phi)\right]
\,.
\end{equation}

The field equations can now be derived in a usual manner.
The differential equation for the scalar is
\begin{equation}
\phi''+(D-1)A'\phi'-V_\phi=0\,,
\end{equation}
where $V_\phi=\frac{dV}{d\phi}$.
The equation for $A$ is written as
\begin{equation}
2[F_x(A'^2)+2A'^2 F_{xx}(A'^2)]A''
-(D-1)
\left[F(A'^2)-2A'^2F_x(A'^2)-\frac{1}{2}(\phi')^2-V(\phi)\right]=0\,,
\end{equation}
where $F_x=\frac{dF(x)}{dx}$ and $F_{xx}=\frac{d^2F(x)}{dx^2}$.
The reparametrization invariance of $y$ leads to the first integral
given by 
\begin{equation}
F(A'^2)-2A'^2F_x(A'^2)+\frac{1}{2}(\phi')^2-V(\phi)=0\,.
\end{equation}

Define $h(x)\equiv 2xF_x(x)-F(x)$. Then, one can find
$h_x(x)=\frac{dh}{dx}=F_x(x)+2xF_{xx}(x)$. Note that
$h(x)$ as well as $F(x)$ does not possess the $x^{D/2}$ term for even $D$.
Using these, the
equations can be rewritten as
\begin{equation}
2\,h_x(A'^2)A''
+(D-1)(\phi')^2=0\,,
\end{equation}
and
\begin{equation}
-h(A'^2)+\frac{1}{2}(\phi')^2-V(\phi)=0\,.
\end{equation}

Now, we take a BPS ansatz \cite{ST,spd1,spd2,spd3,LCS2,LWWZ}
\begin{equation}
A'=-W(\phi)\,.
\end{equation}
Then, the equations reduces to the first order equation
\begin{equation}
\phi'=\frac{2}{D-1}h_x(W^2)W_\phi\,,
\end{equation}
and the potential
\begin{equation}
V(\phi)=\frac{2}{(D-1)^2}[h_x(W^2)]^2 (W_\phi)^2-h(W^2)\,,
\end{equation}
in terms of $W(\phi)$.

If the function $h(x)$ is
a polynomial up to the quadratic order, a domain wall solution exists,
for the potential $V(\phi)$ has two minima
\cite{ST,spd1,spd2,spd3,LCS2,LWWZ}. The solution can be expressed as a
kink solution in the $\phi^4$ theory. We find here that the potential with
many vacua naturally corresponds to the general higher order gravity.

For example, we try to express
the sine-Gordon equation.
We take $W(\phi)=B\phi$, with a positive constant $B$.
Further, we choose
\begin{equation}
h_x(x)=C\,\cos \frac{\alpha\sqrt{x}}{2}\,,
\end{equation}
with constants $\alpha$ and $C$.
This choice is possible if the spacetime dimension $D$ is odd.
Then the scalar obeys the sine-Gordon equation
\begin{equation}
\phi''=-\frac{\alpha B^2C}{D-1}\sin \frac{\alpha B\phi}{2}\,\phi'=
-\frac{\alpha B^3C^2}{(D-1)^2}\sin
\alpha B\phi\,,
\end{equation}
and an exact static solution is known as
\begin{equation}
\phi(y)=\frac{4}{\alpha B}\left(\arctan e^{\frac{\alpha B^2
C}{D-1}y}-\frac{\pi}{4}\right)\,.
\end{equation}
Then, the potential takes the form:
\begin{equation}
V(\phi)=\frac{2B^2C^2}{(D-1)^2}\cos^2 \frac{\alpha
B\phi}{2}-\frac{8C}{\alpha^2}\left(\frac{\alpha B\phi}{2}\sin\frac{\alpha
B\phi}{2}+\cos\frac{\alpha B\phi}{2}\right)+V_0\,.
\end{equation}
The minima of the potential are located at $\alpha B \phi/2=\pi/2\pm
n\pi~(n=0, 1, 2,\dots)$.
Although it is difficult to obtain exact solutions of other types,
we can suppose multiple domain walls with distinct topological numbers
in this and similar models.
The model with many vacua may also serve an interesting mechanism to
realize naturally a small cosmological constant in (thick) brane world. 
This possibility will be studied in future.

%%%%%%%%%%%%%%%%%%%%%%%%%%%%%%%%%%%%%%%%%%%%%%%%%%%%%%%%%%%%%%%%%
%%%%%%%%%%%%%%%%%%%%%%%%%%%%%%%%%%%%%%%%%%%%%%%%%%%%%%%%%%%%%%%%%
\section{Summary and conclusion}
%%%%%%%%%%%%%%%%%%%%%%%%%%%%%%%%%%%%%%%%%%%%%%%%%%%%%%%%%%%%%%%%%
%%%%%%%%%%%%%%%%%%%%%%%%%%%%%%%%%%%%%%%%%%%%%%%%%%%%%%%%%%%%%%%%%
In the present paper, we have attempted to show the possible
quasi-linear second-order theory of gravity
in conformally flat spacetimes.
Models with arbitrary higher order of curvatures have been obtained.
As long as we adopt an isotropic and homogeneous cosmological setting,
the energy conservation holds in the models because there is no scalar
mode and no requirement of frame rescaling.
In spite of them, inflationary expansion can be found in the models.
We have also found that the domain wall solution in the present type of
the higher order gravity can be obtained naturally with the potential
having many minima.

Our work corresponds to the extension of the Lovelock higher-curvature
gravity in arbitrarily higher order terms in higher dimensions.
Our analysis, however, has been limited for the case with the
conformally flat spacetime and the coefficient on the tensorial form of
the action has been still ambiguous.
The stability of the solutions obtained here is problematic for
anisotropic perturbations or tensor modes.
To study it, we should classify the tensorial form of the Lagrangian which
is equvalent only if the Weyl tensor vanishes.
It is known that the dimensionally continued Euler forms have the property
of factorization in terms of those of lower orders when the spacetime is
represented by the direct product of spaces \cite{MH}.
Thus, compactification should be worth studying to seek some special
combination of curvature tensors. It is expected that the hopeful
`critical' relation among the coefficients of different orders of
curvatures will be selected by consideration on various background
spacetimes.

Nevertheless, we emphasize that our arbitrarilty high order gravity can
be applied to many models in various contexts.
The problem of initial singularity can be reconsidered by studying
classical bouncing universes in our model.
On the other hand, the Wheeler-DeWitt equation of the Lagrangian should
lead to higher-derivative quantum cosmology.
The equation must be difficult to treat with, but the study on it may
shed new light on quantum gravity.
The possible black hole solutions are interesting in both cases of
asymptotically flat and asymptotically AdS spacetime.
We shall return to some of the problems in future.

%\acknowledgments
%%%%%%%%%%%%%%%%%%%%%%%%%%%%%%%%%%%%%%%%%%%%%%%%%%%%%%%%%%%%%%%%%%%%%%%%%%%
%Acknowledgements
%%%%%%%%%%%%%%%%%%%%%%%%%%%%%%%%%%%%%%%%%%%%%%%%%%%%%%%%%%%%%%%%%%%%%%%%%%%
%\begin{acknowledgments}
%The authors would like to thank the organizers of JGRG21, where our
%partial result %({\tt [arXiv:10mm.xxxx]}) 
%was presented. %for elucidating comments.
%This study is supported in part by the Grant-in-Aid of Nikaido Research 
%Fund.
%\end{acknowledgments}

%%%%%%%%%%%%%%%%%%%%%%%%%%%%%%%%%%%%%%%%%%%%%%%%%%%%%%%%%%%%%%%%%%%%%%%%%%%
%%%%%%%%%%%%%%%%%%%%%%%%%%%%%%%%%%%%%%%%%%%%%%%%%%%%%%%%%%%%%%%%%%%%%%%%%%%
%%%%%%%%%%%%%%%%%%%%%%%%%%%%%%%%%%%%%%%%%%%%%%%%%%%%%%%%%%%%%%%%%%%%%%%%%%%
%%%%%%%%%%%%%%%%%%%%%%%%%%%%%%%%%%%%%%%%%%%%%%%%%%%%%%%%%%%%%%%%%%%%%%%%%%%

%%%%%%%%%%%%%%%%%%%%%%%%%%%%%%%%%%%%%%%%%%%%%%%%%%%%%%%%%%%%%%%%%%%%%%%%%%%

%%%%%%%%%%%%%%%%%%%%%%%%%%%%%%%%%%%%%%%%%%%%%%%%%%%%%%%%%%%%%%%%%%%%%%%%%%%
%thebibliography
%%%%%%%%%%%%%%%%%%%%%%%%%%%%%%%%%%%%%%%%%%%%%%%%%%%%%%%%%%%%%%%%%%%%%%%%%%%

%\bibliographystyle{apsrev}
\bibliographystyle{apsrev4-1}
%\bibliography{}

\begin{thebibliography}{99}

%01
\bibitem{Schmidt} 
H.-J. Schmidt, {\tt gr-qc/0412038}; {\tt gr-qc/0602017}.

%02
\bibitem{Whitt} 
B. Whitt, Phys. Lett. {\bf B145} (1984) 176.

%03
\bibitem{HL} 
S. W. Hawking and J. C. Luttrel, Nucl. Phys. {\bf B247} (1984) 250.

%04
\bibitem{fR1}
T.~P.~Sotiriou and V.~Faraoni,
%f(R) Theories Of Gravity
Rev. Mod. Phys. {\bf 82} (2010) 45, {\tt arXiv:0805.1726 [gr-qc]}.

%05
\bibitem{fR2}
A. De Felice and S. Tsujikawa,
%f(R) theories
Living Rev. Rel. {\bf 13} (2010) 3.
%arXiv:1002.4928 [gr-qc]

%06
\bibitem{fR3}
S.~Nojiri and S.~D.~Odintsov,
%Unified cosmic history in modified gravity: 
%from F(R) theory to Lorentz non-invariant models
Phys. Rep. {\bf 505} (2011) 59,
{\tt arXiv:1011.0544v4 [gr-qc]}.

%S.~Capozziello and M. De Laurentis,
%%Extended Theories of Gravity.
%Phys. Rep. {\bf 509} (2011) 167,
%{\tt arXiv:1108.6266 [gr-qc]}.


%\bibitem{Nelson}
%W.~Nelson, 
%Phys. Rev. {\bf D82} (2010) 104026,
%%Static Solutions for Fourth Order Gravity
% {\tt arXiv:1010.3986 [gr-qc]};
%Phys. Rev. {\bf D 82} (2010) 124044,
%%Restricting Fourth Order Gravity via Cosmology
% {\tt arXiv:1012.3353 [gr-qc]}.

%S.~Capozziello and A.~Stabile,
%%The Weak Field Limit of Fourth Order Gravity
% {\tt arXiv:1009.3441 [gr-qc]}.

%A.~Stabile and G.~Scelza,
%%Rotation Curves of Galaxies by Fourth Order Gravity
%{\tt arXiv:1107.5351 [gr-qc]}.


%07
\bibitem{Lovelock}
D. Lovelock,
J. Math. Phys. {\bf 12} (1971) 498;
J. Math. Phys. {\bf 13} (1972) 874.


%\bibitem{Briggs}
%C. C. Briggs,
%%Some Possible Features of General Expressions for Lovelock Tensors and
%%for the Coefficients of Lovelock Lagrangians Up to the 15th Order in
%%Curvature (and Beyond)
%gr-qc/9808050;
%%A General Expression for the Quartic Lovelock Tensor
%gr-qc/9703074;
%%A General Expression for the Quintic Lovelock Tensor
%gr-qc/9607033.


%08
\bibitem{JTWheeler}
J. T. Wheeler,
%Symmetric solutions to the Gauss-Bonnet extended Einstein equations
Nucl. Phys. {\bf B268} (1986) 737.

%09
\bibitem{Aragone}
C.~Aragone, 
%geometric stringy gravity
Phys. Lett. {\bf B186} (1987) 151.

%10
\bibitem{MH}
F.~M\"uller-Hoissen, 
%spontaneous compactification with quadratic and cubic curvature terms
Phys. Lett. {\bf B163} (1985) 106;
%Dimensionally Continued Euler Forms, Kaluza-klein Cosmology And
%Dimensional Reduction
Class. Quant. Grav. {\bf 3} (1986) 665.

%11
\bibitem{WP}
B. Whitt,
%Spherically symmetric solutions of general second-order gravity
Phys. Rev. {\bf D38} (1988) 3000.

%\bibitem{Yue}
%R. Yue, D. Zou, T. Yu, P. Li and Z. Yang,
%%Slowly rotating charged black holes in anti-de Sitter third order
%%Lovelock gravity
%Gen. Relativ. Gravit. {\bf 43} (2011) 2103.

%12
\bibitem{Zumino}
B. Zumino,
Phys. Rep. {\bf 137} (1985) 109.

%13
\bibitem{Zwi}
B. Zwiebach,
Phys. Lett. {\bf B156} (1985) 315.

%14
\bibitem{NMG}
E. A. Bergshoeff, O. Hohm and P. K. Townsend,
%Massive Gravity in Three Dimensions
Phys. Rev. Lett. {\bf 102} (2009) 201301.
%arXiv:0901.1766 [pdf, ps, other]

%15
\bibitem{LP}
H. L\"u and C.N. Pope,
%Critical Gravity in Four Dimensions
Phys. Rev. Lett. {\bf 106} (2011) 181302.
%arXiv:1101.1971 [hep-th]

%16
\bibitem{LPP}
H. L\"u, Yi Pang and C.N. Pope,
%Conformal Gravity and Extensions of Critical Gravity
Phys. Rev. {\bf D84} (2011) 064001.
%arXiv:1106.4657 [hep-th]

%17
\bibitem{SGT}
T. \c{C}. \c{S}i\c{s}man, \.{I}. G\"ull\"u and B. Tekin,
%%All unitary cubic curvature gravities in D dimensions
Class. Quant. Grav. {\bf 28} (2011) 195004.
%%arXiv:1103.2307 [hep-th]

%18
\bibitem{Ibison} M. Ibison,
%On the conformal forms of the RW metric
J. Math. Phys. {\bf 48} (2007) 122501.

%19
\bibitem{MO} 
K.~A.~Meissner and M.~Olechowski, 
%Domain Walls without Cosmological Constant in Higher Order Gravity
Phys. Rev. Lett. {\bf 86} (2001) 3708;
%Brane localization of gravity in higher derivative theory
Phys. Rev. {\bf D65} (2002) 064017;

%20
\bibitem{OR}
J. Oliva and S. Ray,
Phys. Rev. {\bf D82} (2010) 124030.

%21
\bibitem{ST}
K. Skenderis and P. K. Townsend,
Phys. Lett. {\bf B468} (1999) 46.

%22
\bibitem{spd1}
I. Low and A. Zee,
%Naked singularity and Gauss-Bonnet term in brane world scenarios
Nucl. Phys. {\bf B585} (2000) 395.

%23
\bibitem{spd2}
Z. Kakushadze,
%Localized (super)gravity and cosmological constant
Nucl. Phys. {\bf B589} (2000) 75.

%24
\bibitem{spd3}
O. Corradini and Z. Kakushadze,
%Naked singularity and Gauss-Bonnet term in brane world scenarios
Phys. Lett. {\bf B494} (2000) 302.

%\bibitem{LCS1}
%H. L. C. Louzada, U. Camara dS and G. M. Sotkov,
%%Massive 3D gravity Big Bounce
%Phys. Lett. {\bf B686} (2010) 268.

%25
\bibitem{LCS2}
U. Camara dS and G. M. Sotkov,
%New massive gravity domain walls
Phys. Lett. {\bf B694} (2010) 94.

%\bibitem{LCS3}
%U. Camara dS, C. P. Constantinidis and G. M. Sotkov,
%%New massive gravity Holography
%arXiv:1009.2665 [hep-th].

%26
\bibitem{LWWZ}
Y.-X. Liu, Y.-Q. Wang, S.-F. Wu and Y. Zhong
%Analytic Solutions of Brane in Critical Gravity
{\tt arXiv:1201.5922 [hep-th]}.

\end{thebibliography}

%%%%%%%%%%%%%%%%%%%%%%%%%%%%%%%%%%%%%%%%%%%%%
%%%%%%%%%%%%%%%%%%%%%%%%%%%%%%%%%%%%%%%%%%%%%
%%%%%%%%%%%%%%%%%%%%%%%%%%%%%%%%%%%%%%%%%%%%%
%%%%%%%%%%%%%%%%%%%%%%%%%%%%%%%%%%%%%%%%%%%%%
%%%%%%%%%%%%%%%%%%%%%%%%%%%%%%%%%%%%%%%%%%%%%
%%%%%%%%%%%%%%%%%%%%%%%%%%%%%%%%%%%%%%%%%%%%%
%%%%%%%%%%%%%%%%%%%%%%%%%%%%%%%%%%%%%%%%%%%%%

%%%%%%%%%%%%%%%%%%%%%%%%%%%%%%%%%%%%%%%%%%%%%%%%%%%%%%%%%%%%%%%%%%%%%%%%%%%
%%%%%%%%%%%%%%%%%%%%%%%%%%%%%%%%%%%%%%%%%%%%%%%%%%%%%%%%%%%%%%%%%%%%%%%%%%%
%%%%%%%%%%%%%%%%%%%%%%%%%%%%%%%%%%%%%%%%%%%%%%%%%%%%%%%%%%%%%%%%%%%%%%%%%%%
%%%%%%%%%%%%%%%%%%%%%%%%%%%%%%%%%%%%%%%%%%%%%%%%%%%%%%%%%%%%%%%%%%%%%%%%%%%
%%%%%%%%%%%%%%%%%%%%%%%%%%%%%%%%%%%%%%%%%%%%%%%%%%%%%%%%%%%%%%%%%%%%%%%%%%%

\end{document}